\documentclass[aps,prl,twocolumn,showpacs,superscriptaddress,groupedaddress]{revtex4} 
\usepackage{graphicx} % needed for figures
\usepackage{dcolumn} % needed for some tables
\usepackage{bm} % for math
\usepackage{amssymb} % for math
\usepackage{amsmath}
\usepackage[tight]{subfigure}

\usepackage[amssymb, textstyle, cdot, binary]{SIunits}
\DeclareMathOperator{\e}{e}						% e de l'exponentielle
\usepackage[babel=true]{csquotes}

\begin{document}

\title{Anisotropic Atom-Surface Interactions in the Casimir-Polder Regime}

\author{T.~Taillandier-Loize} \affiliation{Universit\'{e} Paris 13, Sorbonne Paris Cit\'{e}, Laboratoire de Physique des Lasers, CNRS, (UMR 7538), F-93430, Villetaneuse, France}
\author{J.~Baudon} \affiliation{Universit\'{e} Paris 13, Sorbonne Paris Cit\'{e}, Laboratoire de Physique des Lasers, CNRS, (UMR 7538), F-93430, Villetaneuse, France}
\author{G.~Dutier} \affiliation{Universit\'{e} Paris 13, Sorbonne Paris Cit\'{e}, Laboratoire de Physique des Lasers, CNRS, (UMR 7538), F-93430, Villetaneuse, France}
\author{F.~Perales} \affiliation{Universit\'{e} Paris 13, Sorbonne Paris Cit\'{e}, Laboratoire de Physique des Lasers, CNRS, (UMR 7538), F-93430, Villetaneuse, France}
\author{M. Boustimi} \affiliation{Department of Physics, Umm Al-Qura University, Mekkah, Saudi Arabia} 
\author{M.~Ducloy\footnote{Corresponding author: martial.ducloy@univ-paris13.fr}} \affiliation{Universit\'{e} Paris 13, Sorbonne Paris Cit\'{e}, Laboratoire de Physique des Lasers, CNRS, (UMR 7538), F-93430, Villetaneuse, France}

\date{\today}

\begin{abstract}
The distance-dependence of the anisotropic atom-wall interaction is studied. The central result is the $1/z^6$ quadrupolar anisotropy decay in the retarded Casimir-Polder regime. Analysis of the transition region between non-retarded van der Waals regime (in $1/z^3$) and Casimir-Polder regime shows that the anisotropy cross-over occurs at very short distances from the surface, on the order of $0.03 \lambda$, where $\lambda$ is the atom characteristic wavelength. Possible experimental verifications of this distance dependence are discussed.
\end{abstract}

\pacs{34.35.+a, 03.75.Be, 12.20.Fv}
\maketitle

The force between neutral polarisable systems is a ubiquitous phenomenon in nature, with many applications in physics, chemistry, biology\dots A paramount example is the long-range interaction potential between neutral microscopic quantum systems, like atomic systems, and a solid surface. For plane surfaces this interaction is usually governed by a power-law attractive potential \cite{ref1,ref2}. For atom-surfaces distances $z$ smaller than the wavelengths of the optical transitions involved in the atomic polarisability, the interaction is of the dipole-induced dipole type, and governed by the well-known non-retarded van der Waals potential in $-C_3/z^3$ , which reflects the correlations of dipole fluctuations \cite{ref1}. At larger distances, retardation effects get important, and asymptotically lead to a $-C_4/z^4$ potential, as demonstrated in the pioneering work of Casimir and Polder \cite{ref2}. 

Atom (molecule) - surface forces are central in numerous scientific and technological domains: surface adsorption of atoms, gas-surface equilibrium, cavity QED \cite{ref3}, quantum reflection of atoms on surfaces \cite{ref4}, micro-electromechanical systems \cite{ref5}, research for a fifth fundamental force \cite{ref6}, etc. In most of the above studies, the interaction potential has to be treated in its \textit{full distance} range (retarded and non-retarded), but is generally considered \textit{scalar}. However the atom-surface potential has a cylindrical symmetry around the surface normal, and exhibits a quadrupolar component which may get important for non-scalar energy levels. Anisotropic surface potential strongly alters the internal dynamics and symmetry of nearby atomic systems. For example, surface-induced symmetry break and internal level coupling have been observed on rare-gas metastable states scattered at material surfaces \cite{ref7}. From previous experimental studies of the anisotropic potential, on can underline two points: (i) on one hand, in selective reflection (SR) studies --- generally sensitive to a $\lambda/2\pi$ distance ($\sim \unit{100}{\nano\meter}$) from the surface (see \textit{e.g.} \cite{ref8}) ---, the influence of the anisotropic potential has not been observed, although SR spectroscopy gives access to the excited atomic response \cite{ref9}; (ii) on the other hand, in beam scattering studies, the range of the anisotropic interaction appears to be always smaller than \unit{10}{\nano\meter} --- in general \unit{5}{\nano\meter} or less --- \cite{ref10}. Thus one can state that up to now those anisotropic characteristics have been observed in the non-retarded regime. Indeed retarded interactions, at the lowest order in the asymptotic regime, only depend on the \textit{scalar} static atom polarisability, as first shown by Casimir and Polder \cite{ref2}. To analyse their anisotropic character, one has to go to higher-order components of the atomic response. In this letter, we study the behaviour of the anisotropic response in the retarded regime, and discuss its possible experimental observation in atom-surface scattering.

Many authors have theoretically studied atom-surface interactions \cite{ref11,ref12}. In this work we use a linear response quantum-mechanical approach via field susceptibilities near a plane surface \cite{ref12}. For a dipolar interaction Hamiltonian ($-\bm{D}.\bm{E}$, with $\bm{D}$ the atomic electric dipole operator, and $\bm{E}$ the \textit{e.m}. field), the energy-level shift of an atom in a ground or metastable state, located at position $\bm{r}_0$ can be written under the form of an integral over imaginary frequency $\xi$ (following the approach of Ref~\cite{ref12}):
	\begin{equation}
	\delta E_0 = - \frac{\hbar}{2\pi}\int_0^\infty d \xi G_{\alpha\beta} \left(\bm{r}_0,\bm{r}_0;\imath\xi\right) \alpha_{\alpha\beta}^0\left(\imath\xi\right)
	\label{eq1}
	\end{equation}
where $G_{\alpha\beta}$ and $\alpha_{\alpha\beta}$ are field and atom dipole correlation functions respectively. In Eq.\eqref{eq1}, subscripts $\alpha$ and $\beta$ which denote the Cartesian components ($z$ normal to the surface; $x$, $y$ parallel to the surface) are summed over. For a perfect metallic reflector, the surface-induced field correlation components are given by:
	\begin{align}
	&G_{xx}^S = G_{yy}^S = \left(1+\sigma+\sigma^2\right) \e^{-\sigma}/8z^3 \notag \\
	&G_{zz}^S = \left(1+\sigma\right) \e^{-\sigma}/4z^3 \label{eq2}\\
	&G_{\alpha\beta}^S = 0 \quad \text{for}\quad \alpha\neq\beta \notag
	\end{align}
with $\sigma=2\xi z/c$. For a dielectric reflector, one should multiply~\eqref{eq2} by the surface response $\left(\varepsilon-1\right)/\left(\varepsilon+
1\right)$. The atom polarisability in the ground (metastable) state is given by:
	\begin{equation}
	\alpha_{\alpha\beta}\left(\imath\xi\right)=\frac{2}{\hbar}\sum_n \omega_{0n}\frac{d_\alpha^{0n} \ d_\beta^{0n}}{\omega_{0n}^2+\xi^2}
	\label{eq3}
	\end{equation}
with $d_\alpha^{0n}=\left\langle 0 \left| D_\alpha \right| n \right\rangle$ and $\omega_{0n}$ is the 0-n transition frequency.
In the non-retarded regime ($\omega_{0n} z/c \ll 1$), the integration of Eqé\eqref{eq1} straightforwardly leads to the well-known result
	\begin{equation}
	\delta E_0 = \left\langle 0 \left| H_{vw} \right| 0 \right\rangle \quad\text{with}\quad H_{vw} =- \frac{D_z^2 +\bm{D}^2}{16 \; z^3}
	\label{eq4}
	\end{equation} 				
$H_{vw}$ is the effective van der Waals Hamiltonian in the electrostatic limit. The anisotropic (quadrupolar) interaction potential is given by
	\begin{equation}
	-\frac{Q^{\left(2\right)}}{16z^3} \quad\text{with}\quad Q^{\left(2\right)}= D_z^2-\bm{D}^2/3
	\label{eq5}
	\end{equation}
(irreducible tensorial operator of 2d order)
At the opposite, in the fully retarded limit ($\omega_{0n} z/c \gg 1$), one can use the low-frequency asymptotic limit of the atomic polarisability ($\xi\approx0$) to get the famous Casimir-Polder result:
	\begin{equation}
	\delta E_0^{CP} = -\frac{\hbar c}{8\pi z^4}\left[\alpha_{xx}\left(0\right)+\alpha_{yy}\left(0\right)+\alpha_{zz}\left(0\right)\right]
	\label{eq6}
	\end{equation} 
As discussed above, this retarded interaction involves the scalar polarisability only. To get the quadrupolar component in the asymptotic Casimir-Polder limit, one has to go to higher-order components by expanding $\alpha_{\alpha\beta}$ [Eq.~\eqref{eq3}] over $\xi$: 
	\begin{equation}
	\alpha_{\alpha\beta} \left(\imath\xi\right) \approx \frac{2}{\hbar} \sum_n \frac{d_\alpha^{0n} \ d_\beta^{0n}}{\omega_{0n}}\left(1-\frac{\xi^2}{\omega_{0n}^2}\right)
	\label{eq7}
	\end{equation}
The $\xi^2$ term of Eq.~\eqref{eq7}, when reported in Eq.~\eqref{eq1}, yields a $z^{-6}$ interaction potential with a non-scalar component:
	\begin{equation}
	\delta E_0^R=\frac{c^3}{4\pi z^6} \sum_n \frac{2 \left| d_x^{0n}\right|^2+2 \left| d_y^{0n}\right|^2+  \left| d_z^{0n}\right|^2}{\omega_{0n}^3}
	\label{eq8}
	\end{equation}
the quadrupolar part of which can be written:
	\begin{equation}
	\delta E_{0Q}^R= -\frac{c^3}{4\pi z^6} \sum_n \frac{ \left| d_z^{0n}\right|^2 - \left| d^{0n}\right|^2/3}{\omega_{0n}^3}
	\label{eq9}
	\end{equation}
Let us note that Eq.~\eqref{eq9} can be written under the operational form:
	\begin{equation*}
	\delta E_{0Q}^R= -\frac{\hbar^3 c^3}{4\pi z^6} \left\langle 0 \left| H_{0Q}^R \right| 0 \right\rangle \quad\text{with}\quad 
	\end{equation*}
	\begin{equation}
	H_{0Q}^R = - \frac{1}{3}\bm{D} \frac{1}{\left(H_{at}-\hbar \omega_0\right)^3}\bm{D} + D_z \frac{1}{\left(H_{at}-\hbar \omega_0\right)^3} D_z 
	\label{eq10}
	\end{equation}			 			
$H_{at}$ being the free atom Hamiltonian.
The main property of the anisotropic potential in the Casimir-Polder regime is its $z^{-6}$ dependence, as compared to the scalar Casimir-Polder potential which is in $z^{-4}$. If a virtual dipolar coupling (0-1) is predominant, then Eqs.~\eqref{eq9}-\eqref{eq10} can be written as:
	\begin{equation}
	\delta E_{0Q}^R= -\frac{c^3}{4\pi z^6} \frac{1}{\omega_{01}^3} \left\langle 0 \left| Q^{\left(2\right)} \right| 0 \right\rangle
	\label{eq11}
	\end{equation}

The intermediate atom-wall separation can be analyzed with a full integration of Eq.~\eqref{eq1}, using special functions $f$ and $g$, defined in \cite{ref13}, with the help of sine integral $Si$ and cosine integral $Ci$ functions:
	\begin{align}
	&f\left(t\right) = Ci\left(t\right) \sin t - \left[Si\left(t\right) - \pi/2 \right] \cos t \notag \\
	&g(t) = -Ci\left(t\right) \cos t - \left[Si\left(t\right) - \pi/2 \right] \sin t
	\label{eq12}
	\end{align}
One gets for the quadrupolar interaction:
	\begin{multline}
	\delta E_{0Q} = - \frac{1}{8\pi z^3} \sum_n \left(\left| d_z^{0n}\right|^2-\left| d^{0n}\right|^2/3\right)
	\\
	\left[-z_n+z_n g\left(z_n\right)+\left(1+z_n^2\right) f\left(z_n\right)\right]
	\label{eq13}
	\end{multline}
where $z_n = 2 \omega_{0n} z/c$. One can show that Eq.~\eqref{eq13} can be well approximated by the following analytic expression:
	\begin{equation}
	\delta E_{0Q} \approx -\frac{1}{16 z^3 \left[1+\left(z/\Lambda_2\right)^{5/2}\right]^{6/5}} \left\langle 0 \left| Q^{\left(2\right)} \right|0\right\rangle
	\label{eq14}
	\end{equation}
where $\Lambda_2$ is the cross-over distance. For $z\rightarrow 0$, Eq.~\eqref{eq14} gives the near-field quadrupolar potential. For very large $z$, it yields the fully retarded Casimir-Polder interaction, provided that:
	\begin{equation}
	\Lambda_2^3 = \frac{4c^3}{\pi} \dfrac{\sum\limits_n \dfrac{\left| d_z^{0n}\right|^2-\left| d^{0n}\right|^2/3}{\omega_{0n}^3}}{\sum\limits_n \left| d_z^{0n}\right|^2-\left| d^{0n}\right|^2/3}
	\label{eq15}
	\end{equation}
For a predominant dipolar coupling at frequency $\omega_{01}$ (wavelength $\lambda_{01}$), the cross-over distance is governed by:
	\begin{equation}
	\Lambda_2=\sqrt[3]{\frac{4}{\pi}}\frac{\lambda_{01}}{2\pi}
	\end{equation}
It is worth noting that, making $\Lambda_2 \rightarrow \infty$, one recovers the van der Waals interaction term.

	Up to now, the quadrupolar atom-surface interaction has been mainly explored in the first excited configuration of the rare gases which is characterized by one excited \textit{s}-electron and a core of 5\textit{p}-electrons, $\text{n}p^5 \ \left(\text{n}+1\right)s$. There are two metastable states $\text{J}=0$ and $\text{J}=2$, represented --- in the \textit{L-S} coupling --- as $^3\text{P}_0$ and $^3\text{P}_2$ triplet levels. The anisotropic interaction comes from the $p^5$ core --- and not from the (scalar) \textit{s} electron ---, and increases with the size of the core, \textit{i.e.} the noble gas atomic number. In \textit{L-S} coupling scheme, this anisotropy exists for the $^3\text{P}_{0,1,2}$ states (L = 1), and its symmetry is thus governed by the ($L_z^2-L^2/3$) operator, as expected from the Wigner-Eckart theorem which involves $D_z^3-D^2/3 \propto L_z^2-L^2/3$. It directly links fine-structure changing transitions to surface-induced Zeeman transitions \cite{ref14}. Previous experimental works have been interpreted in the non-retarded van der Waals regime. In the following, we analyze the influence of retardation effects.

The most direct way (presumably not the easiest) to evidence the long distance behaviour of the non scalar component of the interaction is to analyse \textit{inelastic} scattering processes of which this component is specifically responsible. Among such processes, exo-energetic so-called \textit{van der Waals - Zeeman} transitions (\textit{e.g.} transitions from Zeeman state M to state M-1), occurring in the presence of a magnetic field not collinear with the normal to the surface, are good candidates since they have been previously evidenced theoretically and experimentally for metastable rare gas atoms ($^3\text{P}_2$) \cite{ref10,ref15}. In the interpretation of these results, both scalar and non scalar (quadrupolar) parts of the interaction have been assumed of the van der Waals type, \textit{i.e.} proportional to $z^{-3}$. The main difficulty here to move to larger values of $z$ is the very short range (at thermal velocities, a few nm up to a few tens nm) within which such transitions are observable. Larger ranges (\unit{50-100}{\nano\meter}) are expected at lower velocity (a few tens of m/s) experimentally accessible, using a magneto-optical trap from which atoms are pushed away by a laser beam \cite{ref16}. 
If no selection of atom-surface distance $z$ is performed, then both elastic and inelastic diffractions are so largely dominated by short-distance interactions that no retardation effect can be evidenced. Various methods have been successfully used so far to investigate atom-surface interaction in the retarded regime \cite{ref4,ref17,ref18}, but they are restricted to the scalar part of the potential. To better evidence the behaviour of both scalar and non scalar interactions as a function of the distance, the selection of a definite interval of z is clearly needed. Recently, a new method to prepare narrow and non spreading wave packets has been proposed and tested experimentally with slow ($v =\unit{ 50}{\meter\per\second}$) metastable argon atoms Ar*($^3\text{P}_2$) \cite{ref19}. Such so-called \enquote{Michelangelo} wave packets are obtained by passing a well-collimated beam of Ar* atoms (the coherence width of which is of a \unit{few}{\micro\meter}) through a thick standing light wave locked in frequency on the open transition $1s_5$ ($^3\text{P}_2$) - $2p_8$ (J=2), at wavelength $\lambda_\text{op} = \unit{801.5}{\nano\meter}$. Except in the vicinity of the standing wave nodes, atoms are partly transferred to the ground state and then are no longer detectable by standard metastable atom detectors. In fact the best efficiency (\unit{84}{\%}) of this \enquote{optical quenching} is obtained with a circularly polarized light and atoms initially polarized in Zeeman state M = 0. As a consequence, in the following calculations, we shall assume incident atoms in this state. It is shown in \cite{ref19} that the transverse profile of the wave packet is Gaussian, with a width $\delta z$ depending on the laser intensity $I$ as $I^{-1/4}$. As a compromise width \textit{versus} intensity, the following calculations have been carried out at a rather reasonably high laser power, $I = \unit{240}{\milli\watt\per\centi\meter\squared}$, giving $\delta z = \unit{12}{\nano\meter}$. Under such conditions, the Rabi pulsation is $\Omega = 5.28 \Gamma$, where $\Gamma = \unit{$2\pi \times 5.8$}{\mega\hertz}$ is the natural width of the upper level. It is worth noting that, as soon as the wave packet comes out of the light wave, its natural spreading gets back. Then the solid surface under investigation must be placed as close as possible to the light beams. At the velocity considered here, $v = \unit{20}{\meter\per\second}$, within the extension of the surface along axis $x$ (direction of atom trajectories), $\Delta x = \unit{500}{\nano\meter}$, this natural spreading while small (less than $\unit{$\pm 5.8$}{\nano\meter}$) will be taken in account.

Owing to the shortness of the de Broglie atomic wavelength ($\lambda_\text{at} =\unit{0.56}{\nano\meter}$) relatively to distances characteristic of the experiment geometry, the semi-classical approximation is valid. More precisely, the external motion along $z$ is assumed to be classical, with a constant velocity $v$, whereas the internal state is expanded over the Zeeman-state basis set $\left| \text{M}\right\rangle$ where states $\left| \text{M}\right\rangle$ are referred to the direction of the magnetic field $\bm{B}$. The interaction experienced by an atom is then expressed in the form of a $5\times5$ matrix combining the magnetic interaction $W_{MM'} = -\delta_{MM'} M g \mu_B B$ ($g$ is the Landé factor, $\mu_B$ the Bohr magneton) and the interaction with the surface (a planar perfect conductor) described by a $5\times5$ matrix $\underline{\bm{V}}$, whose elements are $V_{MM'}(\alpha, z, t)$, where $\alpha$ is the angle between $\bm{B}$ and the normal to the surface.
 
\begin{figure}[!ht]
	\centering
	\includegraphics[width=0.45\textwidth]{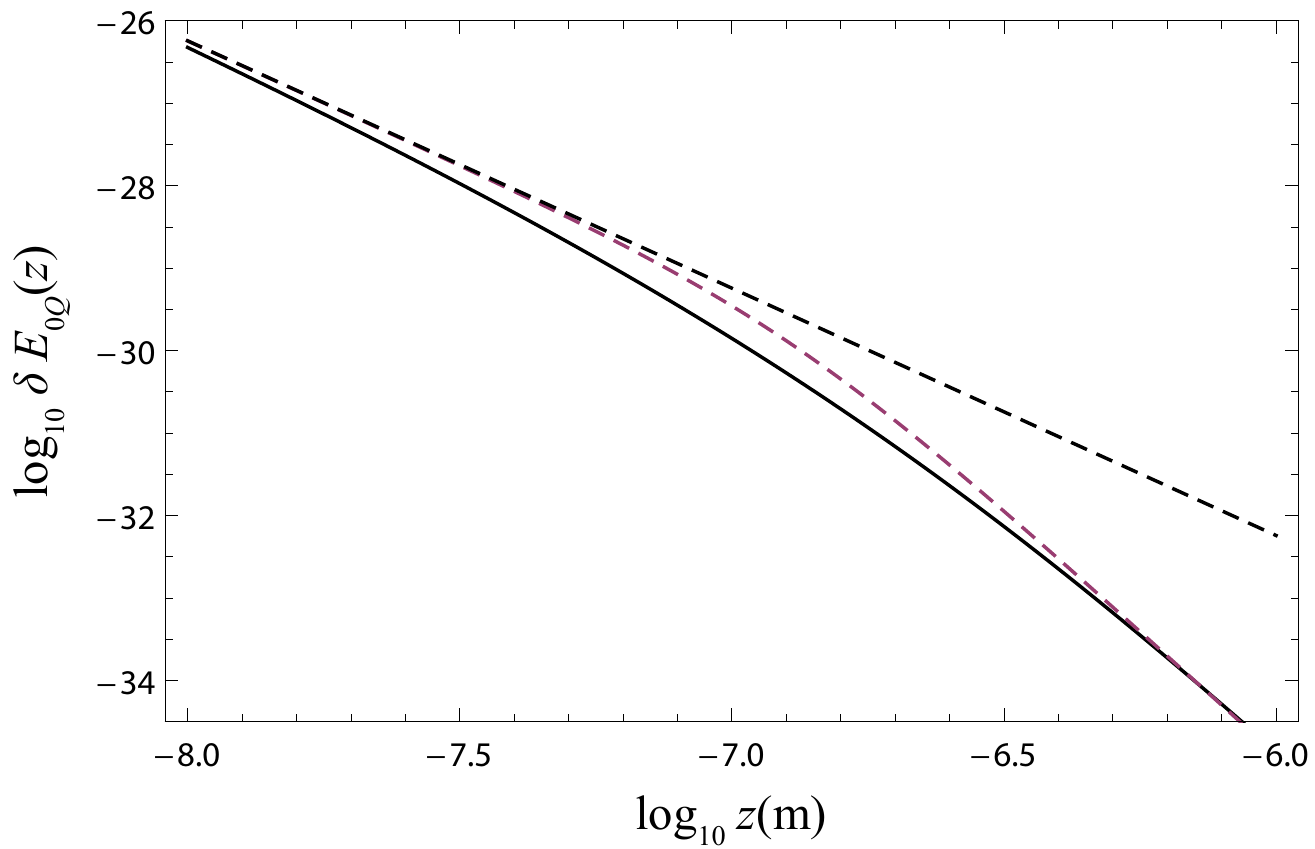}
	\caption{$\log_{10}$ of factors $\delta E_{0Q}\left(z\right)$ involved in the anisotropic part of interaction (see text). Upper dashed line: pure van der Waals interaction; lower curves:  $\delta E_{0Q}^{WS}\left(z\right)$  (full line), $\delta E_{0Q}^{app}\left(z\right)$ (dashed line).}
	\label{fig1}
	\end{figure}

The quadrupolar part of the atom-surface interaction, in the case of a metastable argon atom (Ar* $^3\text{P}_2$) in the vicinity of an almost perfect conductor (gold) takes the form given by Eq.~\eqref{eq13}, the sum being reduced to the dominant $n = 1$ term. This term will be noted $\delta E_{0Q}^{WS}$. As shown previously this expression can be further approximated by (\textit{cf} Eq.\eqref{eq14}):
	\begin{equation}
	\delta E_{0Q}^{app} \left(z \right) \approx - \frac{\eta}{16 z^3 \left[1+\left(z/\Lambda_2 \right)^{5/2} \right]^{6/5}}
	\label{eq17}
	\end{equation}
where the \enquote{quadrupolar constant} is \mbox{$\eta = \unit{-0.15}{\text{atomic units}}$} \cite{ref20}. Using $\lambda_{01} = \unit{815.5}{\nano\meter}$ as the dominant dipolar coupling, one gets the cross-over distance $\Lambda_2 = \unit{140}{\nano\meter}$. Quadrupolar terms $\delta E_{0Q}^{WS}$ and $\delta E_{0Q}^{app}$ are shown in figure~\ref{fig1} together with the latter term at the limit $\Lambda_2$ infinite (non-retarded van der Waals potential). The angular dependence of $\underline{\bm{V}}$ is given by: 
	\begin{equation}
	\underline{\bm{V}} = \delta E_{0Q}\left(z\right) \left[\left(2-3\sin^2 \alpha \right) \underline{\bm{A}} +\sqrt{3/2} \sin \alpha \ \underline{\bm{C}} \left(\alpha\right)\right]
	\label{eq18}
	\end{equation}
where $\underline{\bm{A}}$ is a diagonal constant matrix of elements $\left\lbrace 1, 1/2, 1, 1/2, 1 \right\rbrace$ and $\underline{\bm{C}}$ is a symmetric off-diagonal matrix whose elements are given by:
	\begin{align}
	&C_{21}=C_{-1,-2}=\sqrt{6}\cos\alpha& &C_{20}=C_{0,-2}=\sin\alpha \notag\\
	&C_{2,-1}=C_{1,-2}=C_{2,-2}=0 \\
	&C_{10}=C_{0,-1}=\cos\alpha& &C_{1,-1}=\sqrt{3/2}\sin\alpha \notag
	\end{align}
	
For a planar surface and a homogeneous magnetic field, $\alpha$ is constant ($\alpha = \unit{1.22}{\rad}$ in the following calculations), apart from negligible edge effects (note that for $\alpha = 0$, $\underline{\bm{V}}$ is diagonal and no transition occurs). 

At a given distance $z$ to the surface, the internal state evolution in time is governed by a set of 5 coupled differential equations: 
	\begin{equation}
	\imath \hbar \partial_t a_M =\sum_{M'=-2}^{+2} \left(W_{MM'}+V_{MM'} \right) a_{M'}\left(z,t\right)
	\label{eq20}
	\end{equation}
The initial condition, at time $t_1 = -\Delta x/\left(2v\right)$, is: \mbox{$a_M = \delta_{M0}$}. Solving \eqref{eq20} one gets the final amplitudes at time $t_2 = +\Delta x/\left(2v\right)$. In the following, we shall restrict numerical calculations to the amplitudes $a_0\left(z,t_2\right)$ and $a_{-1}\left(z,t_2\right)$, related respectively to $p_\text{el}\left(z\right)$ probability of elastic scattering and $p_\text{in}\left(z\right)$ probability of the unique inelastic transition $0 \rightarrow -1$ we are interested in \cite{ref21}. The final complex amplitudes $a_0$ and $a_{-1}$ generate, at large distance, scattering amplitudes $F_0$ and $F_{-1}$. In the Fraunhofer regime, at a direction belonging to plane $\left(x,z\right)$ and making an angle $\theta$ with the $x$ axis, one gets (up to a constant multiplicative factor): 
	\begin{multline}
	F_{0,-1} \left(\theta\right) = \int_0^\infty dz \ \rho\left(z-z_m\right) \ a_{0,-1}\left(z,t_2\right)
	 \\
	 \exp\left[-\imath \left( k_{0,-1} \sin \theta + \sqrt{k_{0,-1}^2-k_{0}^2}\ \right) z \right]
	\label{eq21}
	\end{multline}
where $\rho\left(z-z_m\right)$ is the Gaussian amplitude of a Michelangelo wave packet centred at $z_m$, $k_0$ is the initial wave number, $k_{-1} = \left(k_0^2 + g \lvert \mu_B \rvert B/\hbar \right)^{1/2}$ is the wave number in the inelastic channel. In our case, with $k_0 = \unit{$1.124\times 10^{10}$}{\reciprocal\meter}$ and $B = \unit{0.01}{\tesla}$, one gets $k_{-1} = \unit{$1.130\times 10^{10}$}{\reciprocal\meter}$. Let $P_\text{el}\left(z_m\right)$ and $P_\text{in} \left(z_m\right)$ be respectively the integrals over $\theta$ of $\lvert F_0 \rvert^2$ and $\lvert F_{-1} \rvert^2$. As soon as $z_m > \unit{25}{\nano\meter}$, $P_\text{el}$ is very close to 1 ($1-P_\text{el} < 10^{-5}$) and we shall rather consider in more detail the inelastic probability $P_\text{in}$.

\begin{figure}[!ht]
	\centering
	\includegraphics[width=0.45\textwidth]{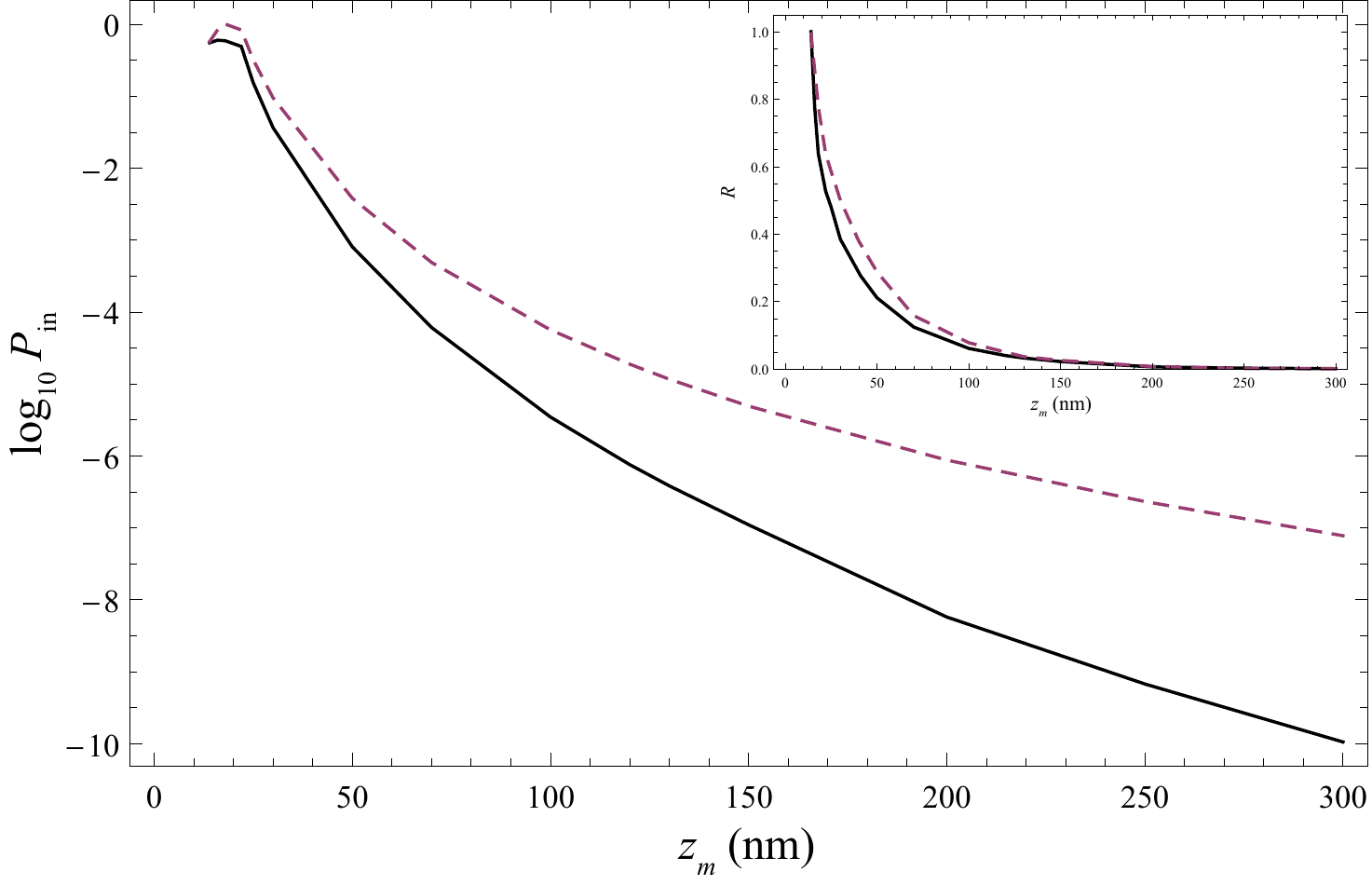}
	\caption{$\log_{10}$ of total inelastic intensities, calculated with van der Waals interactions ($P_\text{in}^{\left(1\right)}$, upper dashed line) and  calculated with retarded potentials ($P_\text{in}^{\left(2\right)}$, lower full line) as functions of the mean distance $z_m$ to the surface. Inset: ratios $R=P_\text{in}^{\left(2\right)}/P_\text{in}^{\left(1\right)}$  (full line) and $R=P_\text{in}^{\left(3\right)}/P_\text{in}^{\left(1\right)}$  (broken line) as functions of $z_m$ ; the cross-over distance is $\Lambda_2 = \unit{140}{\nano\meter}$. }
	\label{fig3}
\end{figure}

Our goal is to compare two kinds of non-diagonal interaction with the surface, namely potentials $V_{MM'}^{\left(1\right)}\left(z,t\right)$ of a pure van der Waals type, leading to $P_\text{in}^{\left(1\right)}\left(z_m\right)$, and retarded potentials defined above, $V_{MM'}^{\left(2\right)}\left(z,t\right)$ issued from $\delta E_{0Q}^{WS}$, $V_{MM'}^{\left(3\right)}\left(z,t\right)$ issued from $\delta E_{0Q}^{app}$, aimed at asymptotically leading to Casimir-Polder interactions at large distance. These latter potentials respectively produce inelastic probabilities $P_\text{in}^{\left(2\right)}\left(z_m\right)$ and $P_\text{in}^{\left(3\right)}\left(z_m\right)$. In figure é\ref{fig3}, $\log_{10}P_\text{in}^{\left(1\right)}$ (upper curve) and $\log_{10}P_\text{in}^{\left(2\right)}$  (lower curve) are plotted as functions of the mean distance $z_m$ to the surface. As expected, at short distances these probabilities are close to each other, whereas at large distances $P_\text{in}^{\left(2\right)}$  decreases much faster than $P_\text{in}^{\left(1\right)}$. These features appear more clearly in inset of figure~\ref{fig3} where the ratios $R = P_\text{in}^{\left(2,3\right)}/ P_\text{in}^{\left(1\right)}$ are plotted as a function of $z_m$. The passage from van der Waals to long-range interactions is readily evidenced by the decrease of the inelastic probability around an amazingly short distance $z_m = \unit{25}{\nano\meter}$, where $R= 0.5$ (while the cross-over distance is $\Lambda_2 = \unit{140}{\nano\meter}$). Up to this distance, $P_\text{in}^{\left(1\right)}$ and $P_\text{in}^{\left(2,3\right)}$ while finally small (of the order of $10^{-5}$) are still measurable. Therefore the long-distance behaviour of the quadrupolar part of the atom-surface interaction should be experimentally investigated, via the observation of inelastic scattering. 

In conclusion, this study of the anisotropic atom-wall potential has put evidence for its very fast decay away from the surface, due to its retarded $z^{-6}$ power-law dependence. The cross-over length of the anisotropic potential amplitude is on the order of $0.05 \lambda$ where $\lambda$ is the characteristic electric dipole wavelength. This explains why, in Selective Reflection studies, up to now, the influence of the surface potential has been mainly limited to isotropic effects. The decay range of inelastic transitions induced by the surface potential is still shorter, on the order of $0.03\lambda$. Its observation in noble gas beam-surface scattering has confirmed this behaviour, showing indeed a practical interaction range equal to or smaller than \unit{5}{\nano\meter} at thermal velocities. The analysis of the retarded anisotropic potential will need slowed atomic beams, in order to increase the atom-surface interaction time and explore the cross-over range between retarded and non-retarded interactions. The atomic motion may thus exhibit some quantum-mechanical features which should be taken into account. This appears to be central in the realisation of novel coherent matter-wave splitters based on anisotropic atom-surface forces \cite{ref22}. The exploration of the anisotropy cross-over range between non-retarded and retarded interactions will have important implications in the interaction between atomic systems and material nanostructures, and the development of hybrid systems \cite{ref23}.


\begin{thebibliography}{99}
\bibitem{ref14}
	However note that configuration mixing also contribute to the anisotropy.
\bibitem{ref21}
	Actually, by solving Eqs (20), all other amplitudes, as well as amplitudes resulting from all other possible initial conditions can be derived.
\end{thebibliography}
\end{document}